\documentclass[12pt]{article}
\usepackage{amsbsy}

\newcommand{\sect}[1]{\setcounter{equation}{0}\section{#1}}

\newcommand{\eq}{\begin{equation}}
\newcommand{\eqa}{\begin{eqnarray}}
\newcommand{\en}{\end{equation}}
\newcommand{\ena}{\end{eqnarray}}
\newcommand{\enn}{\nonumber \end{equation}}

\def\epsihat{{\widehat{\varepsilon}}}
\def\deltahat{ {\widehat\delta} }
\def\Omhat{\widehat{\Om}}

\def\Phihat{\widehat{\Phi}}

\def\sk{\vskip .4cm}
\def\noi{\noindent}
\def\om{\omega}

\def\ga{\gamma}

\let \part\partial

\def\unquarto{{1 \over 4}}

\def\unmezzo{{1 \over 2}}
\def\epsi{\varepsilon}
\def\we{\wedge}

\def\de{\delta}

\def\part{\partial}

\def\sk{\vskip .4cm}

\def\noi{\noindent}

\def\X0{X^0}

\def\om{\omega}

\def\ga{\gamma}

\def\unquarto{{1 \over 4}}
\def\unmezzo{{1 \over 2}}
\def\epsi{\varepsilon}
\def\epsibold{{\bf \epsilon}}

\def\we{\wedge}

\def\de{\delta}

\def\Rhat#1#2{ \Rh^{#1}_{~~~#2} }

\def\Dcal{{\cal D}}
\def\Rcal{{\cal R}}

\def\square{{\,\lower0.9pt\vbox{\hrule \hbox{\vrule height 0.2 cm
\hskip 0.2 cm \vrule height 0.2 cm}\hrule}\,}}

\def\epsilonbar{{\bar \epsilon}}

\def\westar{\we_\star}

\def\omtilde{\tilde \om}
\def\Vtilde{\widetilde{V}}

\def\epsitilde{\widetilde{\epsi}}

\def\phitilde{{\widetilde \phi}}

\def\psibar{\bar \psi}

\def\rhobar{\bar \rho}

\def\zetabar{\bar \zeta}
\def\Om{\Omega}

\def\Sigmabar{\overline \Sigma}

\def\Rhat{\widehat{R}}

\def\Rbold{{\bf R}}
\def\Ombold{{\bf \Om}}
\def\onebold{{\bf 1}}
\def\epsibold{\boldsymbol {\epsilon}}
\def\lambdabold{\boldsymbol {\lambda}}
\def\Phibold{{\bf \Phi}}
\def\Gammabold{{\bf \Gamma}}
\def\Cbold{{\bf C}}

\def\Lfat{{I\!\!L}}

\begin{document}

\begin{titlepage}
\begin{center}
{\Large \bf $OSp(1|4)$ supergravity and its noncommutative extension }
\\[3em]
{\large {\bf Leonardo Castellani} } \\ [2em] {\sl Dipartimento di Scienze e Innovazione Tecnologica
\\ INFN Gruppo collegato di Alessandria,\\Universit\`a del Piemonte Orientale,\\ Viale T. Michel 11,  15121 Alessandria, Italy
}\\ [4em]
\end{center}

\begin{abstract}

\vskip 0.2cm
We review the $OSp(1|4)$-invariant formulation of $N=1$, $D=4$ supergravity and present its noncommutative
extension, based on a $\star$-product originating from an abelian twist with deformation parameter $\theta$. After use of a geometric generalization of the Seiberg-Witten map, we obtain an extended (higher derivative) supergravity theory, invariant under usual $OSp(1|4)$ gauge transformations. Gauge fixing breaks the $OSp(1|4)$ symmetry to its Lorentz subgroup, and
yields a Lorentz invariant extended theory whose classical limit $\theta \rightarrow 0$ is the usual $N=1$, $D=4$ $AdS$ supergravity.

 \end{abstract}

\vskip 8.5cm \noi \hrule \vskip.2cm \noi {\small 
leonardo.castellani@mfn.unipmn.it}

\end{titlepage}

\newpage
\setcounter{page}{1}

\sect{Introduction}

We present a noncommutative (NC) extension of the $OSp(1|4)$-invariant action of $N=1$, $D=4$ anti-De Sitter  supergravity , obtained by the use of a twisted $\star$-product, and  a geometric generalization \cite{AC3} of the Seiberg-Witten map \cite{SW} for abelian twists.  We thus find a higher derivative extension of $OSp(1|4)$ supergravity where the higher order couplings are dictated by the noncommutative structure of the original NC action. The resulting extended theory is geometric (diffeomorphic invariant) and gauge invariant under usual $OSp(1|4)$ gauge transformations.

Noncommutativity of spacetime coordinates
\eq
   [x^\mu , x^\nu] = i \theta^{\mu\nu}  \label{CR}
    \en
is a recurrent theme in physics, being advocated already by Heisenberg in the hope that uncertainty relations between spacetime coordinates 
could  resolve UV divergences arising in quantum field theory \cite{Heis}. 
This motivation still holds, in particular for nonrenormalizable theories of gravity where
 finiteness is the only option for consistency. 
 The issue was explored initially by Snyder in \cite{Snyder}, and
since then noncommutative geometry has found applications in many
branches of physics, mainly in the last two decades. Some
comprehensive reviews can be found in references \cite{Connes},
\cite{Landi}, \cite{madorebook},  \cite{Castellani1}, \cite{revDN},\cite{revSz1}, \cite{book}.

Relations (\ref{CR}) provide a (kinematical) way to encode quantum properties directly in the texture of spacetime. Field theories on noncommuting spacetime can be reformulated as field theories on ordinary (commuting) spacetime, but with a deformed $\star$-product between fields. When the deformation originates from a twist, as in the present paper, the resulting $\star$-product is a twisted product, associative and noncommutative.

This product between fields generates infinitely many derivatives and introduces a
dimensionful noncommutativity parameter $\theta$.  The
prototypical example of twisted product is the
Moyal-Groenewold product \cite{MoyalGroenewold} (historically
arising in phase-space after Weyl quantization \cite{Weyl}) : 
 \eqa & & f(x) \star g(x) \equiv \exp
\left(  {i \over 2} \theta^{\mu\nu}  {\part \over \part x^\mu}
{\part \over \part y^\nu} \right) f(x) g(y) |_{y \rightarrow x} \nonumber \\
& & = f(x) g(x) + {i\over 2} \theta^{\mu \nu}
 \part_\mu f  \part_\nu  g + \cdots + {1 \over n!}  {\left( i \over 2 \right)^n} \theta^{\mu_1\nu_1}
 \cdots \theta^{\mu_n\nu_n} (\part_{\mu_1} \cdots  \part_{\mu_n} f )(\part_{\nu_1} \cdots  \part_{\nu_n} g )+ \cdots \nonumber\\
 & & \label{starproduct}
 \ena
 \noi with a constant $\theta$.  Using this deformed product 
one finds $x^\mu\star x^\nu- x^\nu\star x^\mu=i\theta^{\mu\nu}$, realizing the commutation relations (\ref{CR}).

A straightforward generalization is provided by the twisted $\star$-product, where the partial  derivatives in (\ref{starproduct}) are replaced by a set of commuting
tangent vectors $X_A \equiv  X_A^\mu \part_\mu$.  Dealing with (super)gravity theories, it is desirable to
extend the twisted $\star$-product to forms. This can be done simply by replacing
the tangent vectors $X_A$, acting on functions, with Lie derivatives along $X_A$, acting
on forms. 

Replacing 
products between fields with $\star$-products yields nonlocal actions (called twisted, or NC actions), containing
an infinite number of new interactions and higher derivative terms. In this way 
twisted Yang-Mills theories in flat space have been constructed (see for ex. \cite{Madore,starYM,AJSW}), as well as twisted metric gravity \cite{Wessgroup,book}. Noncommutative $D=4$ vielbein gravity has been treated in \cite{Cham2002,Cham2003}, where
deformations of conformal gravity and complex vielbein gravity were considered, and in \cite{CZ}, where a $U(2,2)$ $\star$-gauge invariant NC action with constraints was proposed as a NC deformation of Einstein gravity. More recently twisted vielbein gravity and its couplings to fermions \cite{AC1}, gauge fields \cite{AC4} and scalars \cite{AC5} have been constructed, as well as a NC deformation of $D=4$, $N=1$ supergravity \cite{AC2}.

These twisted theories are invariant under deformations of the original symmetries. For example the NC action for gauge fields is :
 \eq
 S= {1\over 4 g^2} \int Tr(F_{\mu\nu} \star F^{\mu\nu})
 \en
 where
 \eqa
& & F_{\mu\nu} = \part_{\mu} A_\nu - \part_{\nu} A_\mu - (A_\mu \star A_\nu-
 A_\nu \star A_\mu)\\
 & & A_\mu = A_\mu^I T_I, ~~~~ Tr(T^I T^J)=\de^{IJ}
 \ena
The noncommutative gauge
transformations: 
\eqa 
 & & \de_{\epsi} A_\mu=\part_{\mu} \epsi
-(A_\mu \star  \epsi - \epsi \star  A_\mu)\\ 
 & &\de_{\epsi} F_{\mu\nu} = -(F_{\mu\nu}
\star  \epsi - \epsi \star F_{\mu\nu} ) \ena leave the action invariant, because of the cyclicity of the trace, and 
of the property
 \eq
  \int f \star g = \int g \star f
   \en
 (cyclicity of integral) holding up to boundary terms. 

Noncommutativity apparently comes with a price, i.e. a proliferation of new degrees
of freedom. This can be understood by considering the $\star$-deformation of 
the Yang-Mills field strength:
\eq
F_{\mu\nu}^{I}T_{I} = \part_{\mu} A_{\nu}^{I}T_{I}-\part_{\nu} A_{\mu}^{I}T_{I}-(A_{\mu}^{I} \star A_{\nu}^{J}-
A_{\nu}^{I} \star A_{\mu}^{J})T_{I}T_{J}
\en
Because of noncommutativity of the $\star$-product, anticommutators as well as commutators of
group generators  appear in the right-hand side, and therefore the $T_{I}$ must  be a basis
for the whole universal enveloping algebra of $G$. Thus $I$ runs in principle on the
infinite set of universal enveloping algebra elements (all symmetrized products of the original gauge generators) 
and  the number of independent $A^{I}_{\mu}$ field components increases to infinity. This proliferation 
can be drastically reduced by choosing a specific representation for the generators $T_{I}$. For example if the gauge group is $SU(2)$ and we take its generators to be the in the defining $2 \times 2$ representation, these are just
the Pauli matrices, and a basis for the enveloping algebra only requires an additional matrix proportional 
to the unit matrix.

We may get rid even of these additional degrees of freedom if we use the Seiberg-Witten map, which
allows to express all the fields appearing in the NC action (usually called the NC fields) in terms of  series expansions in $\theta$ containing only the original fields of the undeformed theory, the so called classical fields. The map is engineered so that the classical gauge transformations on the classical fields induce the NC gauge transformations on the NC fields. 
In the $SU(2)$ example, the map relates the four noncommutative fields to 
the three classical $SU(2)$ gauge fields.  

Substituting in the action the NC fields with their expressions in terms of the classical fields yields an infinite series in powers of $\theta$, whose 0-th order term is the classical action.  This higher derivative action is {\it invariant under the
classical gauge variations}, since these by construction induce the NC symmetries of the NC action. Every higher order term in the $\theta$ expansion is actually separately invariant, because the classical symmetries do not involve $\theta$.

With this procedure the NC deformation of vielbein gravity, found in \cite{AC1}, has been re-expressed in \cite{AC3} in terms of the classical vielbein and spin connection, and its Lorentz invariant (and higher derivative) geometric action has been computed up to second order in the noncommutativity parameter \cite{ACD2}. The Seiberg-Witten (SW) map was also used in 
\cite{CZ} to compute the first order correction of the deformed $U(2,2)$ gauge invariant and constrained theory, and
in \cite{Top1,Top2,Top3} for the Mac Dowell-Mansouri gauge theory of gravity.
We also mention the NC extension of $SO(2,3)$ $AdS$ gravity of ref. \cite{DRS}, which contains its expansion to order 
$\theta^2$, and ref. \cite{DGEFV} where the SW map for pure gravity is examined at second order. 

In the present paper we apply this method to $OSp(1|4)$ supergravity. For reviews on the $OSp(1|4)$ formulation of supergravity see for example  \cite{CDF,PvN1,PvN2}. The classical theory contains 
the vielbein $V^a$, the spin connection $\om^{ab}$, the gravitino $\psi$ and nondynamical auxiliary fields (a scalar, a pseudoscalar, a vector and a spin 1/2 fermion) necessary to ensure the full off-shell invariance (and closure) under local $OSp(1|4)$ gauge transformations. The auxiliary fields satisfy 
$OSp(1|4)$-invariant constraints. The $OSp(1|4)$ symmetry
can be exploited to reach a gauge (the soldering gauge) in which the auxiliary fields take constant values. This gauge choice breaks 
the supergroup $OSp(1|4)$ to its Lorentz $SO(1,3)$ subgroup, and reproduces the Mac Dowell-Mansouri action \cite{MDM}, equivalent up to boundary terms to the action of usual $N=1$, $D=4$ anti De Sitter supergravity. For this action supersymmetry is not a {\it gauge} symmetry any more, since it gets broken along with the translations (the $SO(2,3)$ boosts).  However supersymmetry is still ``alive" in the gauge fixed theory. This can be seen in two distinct ways:

i) by solving the supertorsion constraint and passing to second order formalism (expressing the spin connection in terms of the vielbein and the gravitino fields) \cite{FFvN};

ii) or, remaining in first order formalism, by an appropriate modification of the spin connection  supersymmetry variation \cite{DZ}.

Note that the supersymmetry transformations leaving 
the gauge fixed action invariant do not close off-shell (whereas the
$OSp(4|1)$ gauge variations close off-shell by construction).

After $\star$-deforming the product in the $OSp(1|4)$ supergravity action, and using the geometric Seiberg-Witten map , the resulting higher derivative theory contains the same fields as the classical  theory, and is invariant under the same local $OSp(1|4)$ symmetries. 

The reason we start from the $OSp(1|4)$ gauge-invariant theory resides in that all local symmetries (except general coordinate invariance) are contained in a gauge supergroup. The derivation of the Seiberg-Witten map in \cite{SW} is purely algebraic, and nothing changes in the derivation if groups are replaced by supergroups, connections by superconnections etc. In the present paper we apply the map to
$OSp(1|4)$ superconnections and supermatrix  (adjoint) auxiliary fields, containing all the fields of $N=1$, $D=4$ supergravity.  Thus we are guaranteed that supersymmetry (part of the $OSp(1|4)$ symmetry) survives in the extended theory. 

By choosing the same gauge as in the classical theory (the gauge group $OSp(1|4)$ is the same), we obtain an extended theory containing only the vielbein, spin connection and gravitino fields, reducing in the commutative limit to  $N=1$, $D=4$ $AdS$ supergravity. 

The ``mother", non gauge-fixed extended theory is $OSp(1|4)$-invariant, and as such is a locally supersymmetric higher derivative theory. The price to pay for realizing this local gauge supersymmetry (closing off-shell) is the presence of constrained auxiliary fields.

The plan of the paper is as follows. In Section 2 we briefly review  $OSp(1|4)$ supergravity. In Section 3 we recall its manifestly 
$OSp(1|4)$-invariant action. The noncommutative deformation is presented in Section 4. Section 5 deals with  the geometric Seiberg-Witten map, applied in Section 6 to obtain the extended $OSp(1|4)$ supergravity action to second order in $\theta$. Section 7 contains some conclusions.

\sect{Classical $OSp(1|4)$ supergravity}

\subsection{Geometric MacDowell - Mansouri action}

The Mac Dowell-Mansouri action \cite{MDM} for $N=1$, $D=4$ supergravity can be recast in an index-free form:
\eq
 S = 2i \int Tr( R \we R \ga_5 + 2 \Sigma \we \Sigmabar \ga_5) \label{action1}
  \en
  where the trace is taken on spinor indices, and the $2$-form curvatures $R$ (bosonic) and $\Sigma$ (fermionic) originate from the $1$-form $OSp(1|4)$ connection supermatrix:
 \eq
  \Ombold \equiv 
\left(
\begin{array}{cc}
  \Om &  \psi    \\
 \psibar  &  0   \\
\end{array}
\right), ~~~ \Om \equiv \unquarto \om^{ab} \ga_{ab} - {i \over 2} V^a \ga_a
    \label{Omdef}  
  \en
 whose corresponding $OSp(1|4)$ curvature supermatrix is
 \eq
      \Rbold =  d \Ombold - \Ombold \we \Ombold~
  \equiv  \left(
\begin{array}{cc}
   R &  \Sigma    \\
 \Sigmabar  &  0   \\
\end{array}
\right) \label{Rdef}
       \en
  Immediate matrix algebra yields\footnote{we omit wedge products between forms, and all index contractions involve the Minkowski metric $\eta_{ab}$}:
   \eqa
    & & R = \unquarto R^{ab} \ga_{ab} - {i \over 2} R^a \ga_a  \label{defR}\\
    & & \Sigma = d \psi - \unquarto \om^{ab} \ga_{ab} \psi + {i \over 2} V^a \ga_a \psi \label{defSigma} \\
    & & \Sigmabar = d \psibar - \unquarto \psibar \om^{ab} \ga_{ab} + {i \over 2}  \psibar V^a \ga_a 
    \ena
    \noi with 
     \eqa
     & & R^{ab} \equiv d \om^{ab} - \om^{ac} \om^{cb} + V^a V^b + {1 \over 2} \psibar \ga^{ab} \psi \\
     & & R^{a} \equiv d V^{a} - \om^{ab} V^{b}  -  {i \over 2} \psibar \ga^a \psi 
     \ena
 \noi We have also used the Fierz identity for $1$-form Majorana spinors:
  \eq
  \psi \psibar = \unquarto ( \psibar \ga^a \psi \ga_a - \unmezzo \psibar \ga^{ab} \psi \ga_{ab} )
  \en
\noi (to prove it, just multiply both sides by $\ga_c$ or $\ga_{cd}$ and take the  trace on spinor indices).  The $1$-forms $V^a$, $\om^{ab}$ and $\psi$ are respectively the vielbein, the spin connection and the gravitino field (a Majorana spinor, i.e. $\psibar = \psi^T C$, where $C$ is the charge conjugation matrix).

 Carrying out  the spinor trace in  the action (\ref{action1}) yields the familiar MacDowell-Mansouri action:
  \eq
   S = 2 \int \unquarto R^{ab} \we R^{cd} \epsi_{abcd} - 2i \Sigmabar \we \ga_5 \Sigma 
   \en
  After inserting the curvature definitions the action takes the form
   \eq
    S = \int \Rcal^{ab} V^c V^d \epsi_{abcd} + 4 \rhobar \ga_a \ga_5 \psi V^a + \unmezzo (V^a V^b V^c V^d + 2 \psibar \ga^{ab} \psi V^c V^d ) \epsilon_{abcd}  \label{adsSG}
    \en
  with
   \eq
    \Rcal^{ab} \equiv d \om^{ab} - \om^{ac} \om^{cb} , ~~\rho \equiv d \psi - \unquarto \om^{ab} \ga_{ab} \psi \equiv \Dcal \psi
     \label{RLordef}
     \en
   We have dropped the topological term $\Rcal^{ab} \Rcal^{cd} \epsilon_{abcd}$ (Euler form) and used 
   the gravitino Bianchi identity 
    \eq
     \Dcal \rho = - \unquarto \Rcal^{ab} \ga_{ab}
   \en
   and the gamma matrix identity $2 \ga_{ab} \ga_5 = i \epsilon_{abcd} \ga^{cd}$  to recognize that $\unmezzo \Rcal^{ab} \psibar \ga^{cd} \psi \epsilon_{abcd} - 4 i \rhobar \ga_5 \rho$
    is a total derivative. Bianchi identities are easily obtained by taking the exterior derivative of the curvature definitions
    in (\ref{Rdef}), or in (\ref{RLordef}). The action (\ref{adsSG}) describes $N=1$, $D=4$ anti-De Sitter supergravity,
    the last term being the supersymmetric cosmological term. After rescaling the vielbein and the gravitino as
    $V^a \rightarrow \lambda V^a$, $\psi  \rightarrow \sqrt{\lambda} \psi$ and dividing the action by $\lambda^2$, the usual (Minkowski)
    $N=1$, $D=4$ supergravity is retrieved by taking the limit $\lambda \rightarrow 0$. This corresponds to the
Inon\"u-Wigner contraction of $OSp(1|4)$ to the superPoincar\'e group.
 
   The action (\ref{action1}) can be rewritten even more compactly using  the $OSp(1|4)$ curvature supermatrix $\Rbold$:
    \eq
     S =  4 \int {\rm STr} ( \Rbold (\onebold + {\Gammabold^2 \over 2}) \Rbold \Gammabold)    \label{action2}
    \en
    where STr is the supertrace and $\Gammabold$ is the following constant matrix:
     \eq
 \Gammabold \equiv  \left(
\begin{array}{cc}
  i \ga_5 &  0    \\
 0 &  0   \\
\end{array}
\right) \label{Gammabold}
       \en
\noi All boldface quantities are  5 $\times$ 5 supermatrices. 
       
 \subsection{$OSp(1|4)$ gauge variations}
 
 The gauge transformation of the  connection $\Ombold$ 
  \eq
   \de_{\epsibold} \Ombold = d \epsibold - \Ombold \epsibold + \epsibold \Ombold
       \en
       where $\epsibold$ is the $OSp(1|4)$ gauge parameter:
        \eq
         \epsibold \equiv 
         \left(
\begin{array}{cc}
 \unquarto  \epsi^{ab} \ga_{ab} - {i \over 2} \epsi^a \ga_a &  \epsilon   \\
 \epsilonbar  &  0   \\
\end{array}
\right)   \label{gaugeparam}
 \en
 becomes, on the component fields entering $\Ombold$:
 \eqa
  & & 
  \de \om^{ab} = d \epsi^{ab} - \om^{ac} \epsi^{cb} + \om^{bc} \epsi^{ca}- \epsi^a V^b + \epsi^b V^a - \epsilonbar \ga^{ab} \psi  \\
  & & \de V^a = d \epsi^a - \om^{ab} \epsi^b + \epsi^{ab} V^b + i \epsilonbar \ga^a \psi \\  
  & & \de \psi = d \epsilon - \unquarto \om^{ab} \ga_{ab} \epsilon + {i \over 2} V^a \ga_a \epsilon + \unquarto \epsi^{ab} \ga_{ab} \psi - {i \over 2} \epsi^a \ga_a \psi \label{gaugetransf}
   \ena
   Similarly from the gauge variation of the curvature $\Rbold$:
    \eq
    \de_{\epsibold} \Rbold =  - \Rbold \epsibold + \epsibold \Rbold   \label{Rgauge}
     \en
   we find the gauge transformations of the curvature components:
    \eqa
    & &   \de R^{ab} = - R^{ac} \epsi^{cb} + R^{bc} \epsi^{ca}- \epsi^a R^b + \epsi^b R^a - \epsilonbar \ga^{ab} \Sigma \\
    & & \de R^a = - R^{ab} \epsi^b + \epsi^{ab} R^b + i \epsilonbar \ga^a \Sigma \\  
  & & \de \Sigma =  - \unquarto R^{ab} \ga_{ab} \epsilon + {i \over 2} R^a \ga_a \epsilon + \unquarto \epsi^{ab} \ga_{ab} \Sigma - {i \over 2} \epsi^a \ga_a \Sigma
  \ena
  As is well known, the action (\ref{action2}), although a bilinear in the $OSp(1|4)$ curvature, is {\it not} invariant under the 
  $OSp(1|4)$ gauge transformations. In fact it is not a Yang-Mills action (involving the exterior product of $\Rbold$ with its Hodge dual), nor a topological action of the form $\int \Rbold \Rbold$: the constant supermatrix $\Gammabold$ ruins the $OSp(1|4)$ gauge invariance, and breaks it to its Lorentz subgroup. This can be seen easily by noting that the gauge parameter in (\ref{gaugeparam}) commutes with $\Gammabold$ only when restricted to Lorentz rotations ($\epsi^{a} = \epsilon =0$), so that Lorentz rotations indeed leave the action invariant since the supertrace is cyclic. On the other hand a gauge parameter supermatrix containing also translation and/or supersymmetry parameters does not commute with $\Gammabold$, and therefore the action is not invariant under $OSp(1|4)$ translations or supersymmetry transformations. 
  
However supersymmetry is still there: to see it one needs to modify the $\om^{ab}$ supersymmetry transformation. 

\subsection{Supersymmetry}

The (non vanishing) variation of the action (\ref{action2}) under gauge supersymmetry can be computed rather quickly by using $\de \Rbold = [\epsibold , \Rbold]$ with $\epsibold$ containing only the off-diagonal fermionic supersymmetry parameter $\epsilon$. The result is
 \eq
 \de S = -4 \int R^a \rhobar \ga_a \ga_5 \epsilon
 \en
  Now consider instead the variation of the action under an {\it arbitrary} variation of the spin connection $\om^{ab}$, i.e. the variation that defines the $\om^{ab}$ field equation. To compute it with a minimum of algebra, first vary (\ref{action2}) with respect to $\Ombold$, and then set $\de V^a = \de \psi =0$ in $\de \Ombold$ as defined by (\ref{Omdef}). The result is
 \eq
 \de S = 16 \int R^a V^b \de \om^{cd} \epsilon_{abcd}
 \en
Requesting this variation to vanish for arbitrary $\de\om^{ab}$ yields the spin connection
field equation $R^a =0$.

Thus if we consider a  supersymmetry variation of the action, where the variation of $\om^{ab}$ is modified by an extra piece (in addition to its gauge variation):
  \eq
 \de \om^{ab} = \de_{gauge} \om^{ab} + \de_{extra} \om^{ab}
  \en
  the corresponding variation of the action (\ref{action2}) will be
 \eq
 \de S = - 4 \int R^a ( \rhobar \ga_a \ga_5 \epsilon - 2 \de_{extra} \om^{bc} V^d \epsilon_{abcd})
  \en
 This variation can be made to vanish in two distinct ways:
 \sk
\noi 1) by enforcing the constraint $R^a = 0$, which is really equivalent to the field equation of $\om^{ab}$. As is well known $R^a =0$ allows to express the spin connection $\om^{ab}$ in terms of the vielbein and gravitino fields. Substituting back $\om^{ab} (V, \psi)$ in the action leads to the supersymmetric action of AdS supergravity in second order formalism. In this formalism one never needs to vary the fields inside the ``package" $\om^{ab} (V, \psi)$, since any variation of $S$ due to $\de \om^{ab}$ vanishes identically, being proportional to $R^a$ (then one works in the so-called  ``1.5 order formalism"). 
  \sk
  \noi 2) by choosing $\de_{extra} \om^{ab}$ so that
  \eq
   \rhobar \ga_a \ga_5 \epsilon - 2 \de_{extra} \om^{bc} V^d \epsilon_{abcd}=0
  \en
 This equation can be solved for $\de_{extra} \om^{ab}$ in the same way one solves $R^a =0$  for $\om^{ab}$. The result is
 \eq
 \de_{extra} \om^{ab}  =  {1 \over 2} \epsilon^{abcd} (\rhobar_{de} \ga_c \ga_5 \epsilon + \rhobar_{ec} \ga_d \ga_5 \epsilon - \rhobar_{cd} \ga_e \ga_5 \epsilon) V^e
 \en
\noi where $\rhobar_{cd}$ are the components along the vielbein basis of   the gravitino curvature, i.e. $\rhobar \equiv \rhobar_{cd} V^c V^d$.

Thus the first order action (\ref{action2}) is invariant under the supersymmetry transformations, given by eq.s (\ref{gaugetransf}) for the vielbein and the gravitino:
 \eq
  \de V^a = -i \epsilonbar \ga^a \psi, ~~\de \psi = d \epsilon - \unquarto \om^{ab} \ga_{ab} \epsilon \equiv \Dcal \epsilon
  \en
and by the modified rule for $\om^{ab}$:
 \eq
  \de \om^{ab} = \de_{gauge} \om^{ab}+ \de_{extra} \om^{ab}= - \epsilonbar \ga^{ab} \psi  +  {1 \over 2} \epsilon^{abcd} (\rhobar_{de} \ga_c \ga_5 \epsilon + \rhobar_{ec} \ga_d \ga_5 \epsilon - \rhobar_{cd} \ga_e \ga_5 \epsilon) V^e
 \en
 More details can be found for ex. in \cite{PvN1,PvN2}.

\sect{The manifestly $OSp(1|4)$-invariant action}

 Can we reformulate supergravity in an explicit $OSp(1|4)$-invariant way? The answer is yes \cite{Cham1,Cham2,DDFR,PV}, and generalizes the $SO(2,3)$ formulation of AdS gravity of ref.s \cite{West,Fre,StelleWest,DDFR}. Indeed looking at (\ref{action2}), we see that promoting the constant matrix $\Gammabold$ to a field supermatrix $\Phibold$
transforming under $OSp(1|4)$ as
  \eq
   \de \Phibold = - \Phibold \epsibold + \epsibold \Phibold
  \en
 the action $S$ becomes:
   \eq
     S =  \int {\rm STr} ( \Rbold (\onebold + {\Phibold^2 \over 2}) \Rbold \Phibold)    \label{action3}
    \en
and is manifestly $OSp(1|4)$-invariant. By doing so, we are introducing new, auxiliary fields contained in $\Phibold$. We have to ensure, however, that a particular gauge choice exists such that $\Phibold$ reduces to the constant supermatrix $\Gammabold$: only if this gauge choice exists the theory is equivalent to the one described by (\ref{action2}). To satisfy this requirement we choose $\Phibold$ in the symmetric (traceless) 5 -dimensional representation of $OSp(1|4)$  \cite{Cham1}:
     \eq
 \Phibold (x) \equiv  \left( 
\begin{array}{cc}
   {1 \over 4} \pi (x) + i \phi (x) \ga_5 +  \phi^a (x) \ga_a \ga_5  &  \zeta  (x)  \\
 -\zetabar (x) &  \pi (x)    \\
\end{array}
\right) \label{Phidef}
       \en
  Now translations and supersymmetries of  $OSp(1|4)$ can be used to set $\phi^a$ and $\zeta$ to zero \cite{Cham1}.
Moreover,  the $OSp(1|4)$-invariant constraint
 \eq
  \Phibold ^3 + \Phibold = 0 \label{constraint}
  \en
  enforces $\pi =0$ and $\phi = \pm 1$, reducing $\Phibold$ to the constant supermatrix $ \pm \Gammabold$ (ignoring the trivial solution $\Phibold = 0$. If we want to exclude it, we can instead impose the $OSp(1|4)$-invariant constraints $ {\rm STr}(\Phibold^2) =  4 (const)^2$, ${\rm STr}(\Phibold^3)=0$ see ref. \cite{Cham2}).
 The simplest way to implement the constraint  (\ref{constraint}) is to add a ($OSp(1|4)$-invariant) Lagrange multiplier term in the action:
 \eq
  S_{\lambdabold} =  \int {\rm STr} ( \lambdabold \Phibold (\Phibold^2 + \onebold) \Phibold ~D \Phibold D \Phibold D \Phibold D \Phibold )
  \en
 where the Lagrange multiplier $\lambdabold (x)$ is proportional to the unit matrix, i.e. $\lambdabold (x) = \lambda (x) \onebold$, generalizing the analogous term in the $SO(2,3)$-invariant formulation of gravity (see for ex. \cite{StelleWest,Fre}).

Another interesting possibility is to give dynamics (cf. \cite{Cham1}) to the fields $\pi(x)$ and $\phi(x)$ with a potential admitting a stable minimum for the values $\pi =0$ and $\phi = const$. In this paper the constrained auxiliary fields 
are considered as {\it background} fields, on the same footing of the background vector fields that define the 
$\star$-product (see next Section). We do not introduce Higgs fields to break spontaneously the $OSp(1|4)$
invariance. The breaking of $OSp(1|4)$, and contact with $AdS$ $D=4$ supergravity, is made by explicit gauge fixing.

 The $OSp(1|4)$ gauge invariant formulation of $N=1$, $D=4$ anti De Sitter supergravity is our starting point for a noncommutative supersymmetric extension. 

\sect{Noncommutative $OSp(1|4)$ supergravity}

\subsection{The NC action}
The NC theory is obtained by a $\star$-deformation of the action in (\ref{action3}):
 \eq
     S =  \int {\rm STr} ( \Rbold \star (\onebold + {\Phibold \star \Phibold \over 2}) \we_{\star} \Rbold  \star \Phibold)    \label{action4}
    \en
  where the curvature $2$-form $\Rbold$ is now:
   \eq
     \Rbold = d \Ombold - \Ombold \we_\star \Ombold \label{Rstar}
      \en
   and the  $\star$-exterior product between forms is defined as 
 \eqa
 & &  \tau \westar \tau' \equiv  \sum_{n=0}^\infty \left({i \over 2}\right)^n \theta^{A_1B_1} \cdots \theta^{A_nB_n}
   (\ell_{A_1} \cdots \ell_{{A_n}} \tau) \we  (\ell_{{B_1}} \cdots \ell_{{B_n}} \tau')  \nonumber \\
  & & ~~ = \tau \we \tau' + {i \over 2} \theta^{AB} (\ell_{A} \tau) \we (\ell_{B} \tau') + {1 \over 2!}  {\left( i \over 2 \right)^2} \theta^{A_1B_1} \theta^{A_2B_2}  (\ell_{{A_1}} \ell_{{A_2}} \tau) \we
 (\ell_{{B_1}} \ell_{{B_2}} \tau') + \cdots \nonumber \\
  \label{defwestar}
\ena
       \noi where $\ell_{A}$ are Lie derivatives along commuting
       vector fields $X_A$. This noncommutative product is associative due to $[X_A , X_B]=0$.  If  the vector fields $X_A$ are chosen to  coincide with the partial derivatives
$\partial_\mu$, and if $\tau$, $\tau'$ are $0$-forms, then $\tau\star\tau'$ reduces to the well-known Moyal-Groenewold product
\cite{MoyalGroenewold}. 
            
       The $\star$-gauge transformations of the NC fields are: 
   \eqa
    & &  \de_{\epsibold} \Ombold =  d \epsibold - \Ombold \star \epsibold + \epsibold \star \Ombold   \label{Omstargauge} \\
      & &  \de_{\epsibold} \Phibold =  - \Phibold \star \epsibold + \epsibold \star \Phibold   \label{Phistargauge} 
       \ena
      Recalling  the $\star$-gauge transformation of the curvature induced by (\ref{Omstargauge}):
      \eq
    \de_{\epsibold} \Rbold =  - \Rbold \star \epsibold + \epsibold \star \Rbold   \label{Rstargauge}
     \en
     and the cyclicity of the supertrace and of the integral\footnote{twisted differential geometry is treated for ex. in \cite{book}; see the Appendix of \cite{AC2} for a summary.} , the action (\ref{action4}) is manifestly invariant under the $\star$-gauge symmetry.

   Because of noncommutativity, the $\star$-symmetry group is enhanced to $U(1,3|1)$ so as to contain all enveloping algebra generators. Thus the NC $1$-form connection is given by
 \eq
  \Ombold =
\left(
\begin{array}{cc}
  \Om &  \psi    \\
 \psibar  &  w   \\
\end{array}
\right), ~~~ \Om \equiv \unquarto \om^{ab} \ga_{ab}  + i \om I + \omtilde \ga_5 - {i \over 2} V^a \ga_a
- {i \over 2} \Vtilde^a \ga_a \ga_5
    \label{NCOmdef}   \en
 and correspondingly the gauge parameter supermatrix $\epsibold$ becomes
 \eq
  \epsibold =
\left(
\begin{array}{cc}
  \epsi &  \epsilon   \\
 \epsilonbar  &  \eta   \\
\end{array}
\right), ~~~ \epsilon \equiv \unquarto \epsi^{ab} \ga_{ab}  + i \epsi I + \epsitilde \ga_5 - {i \over 2} \epsi^a \ga_a
- {i \over 2} \epsitilde^a \ga_a \ga_5
    \label{NCepsidef}   \en
 containing all the gauge parameters of the superalgebra $U(1,3|1)$.  

The curvature supermatrix $\Rbold$, 
\eq
      \Rbold 
  \equiv  \left(
\begin{array}{cc}
   R &  \Sigma    \\
 \Sigmabar  &  r   \\
\end{array}
\right)
       \en
\noi defined in (\ref{Rstar}), is now given by 
\eqa
 & & R = d \Om - \Om \westar \Om - \psi \westar \psibar \\
 & & \Sigma = d \psi - \Om \westar \psi - \psi \westar w \\
 & & \Sigmabar = d \psibar - \psibar \westar \Om - w \westar \psibar \\
 & & r = dw - \psibar \westar \psi - w \westar w
 \ena 
\noi where $R$ has components along the complete Dirac basis.

As usual in NC theories, the algebra of gauge transformations
 closes as follows:
  \eq
    [\de_{\epsibold_1},  \de_{\epsibold_2}] = \de_{\epsibold_1 \star \epsibold_2 - \epsibold_2 \star \epsibold_1}
     \en
 
 Consistency with the $\star$-gauge transformations requires for the $0$-form $\Phibold$ a similar expansion:
   \eq
  \Phibold =
\left(
\begin{array}{cc}
  \Phi &  \zeta   \\
 -\zetabar  &   \pi   \\
\end{array}
\right), ~~~ \Phi \equiv {i \over 4} \phi^{ab} \ga_{ab}  + {1\over 4} \pi I + i\phi \ga_5 +  \phi^a \ga_a
+   \phitilde^a \ga_a \ga_5
    \label{NCPhidef}   \en
 The crucial difference between the two supermatrix fields $\Ombold$ and $\Phibold$ (besides their different form degree) is their
commutative  limit. We will see in Section 6 how the Seiberg-Witten map ensures that, in the $\theta \rightarrow 0$ limit,
$\Ombold$ contains only $C$-antisymmetric gamma matrices (cf. (\ref{Omdef})) and  $\Phibold$ only  $C$-symmetric gamma matrices 
(cf. (\ref{Phidef})). 

In analogy with the classical case we  also require the $U(1,3|1)$-invariant constraint:
 \eq
  \Phibold \star \Phibold \star \Phibold + \Phibold = 0 \label{constraint2}
   \en
   reducing to (\ref{constraint}) for $\theta \rightarrow 0$. In alternative, we can require ${\rm STr} (\Phibold \star \Phibold) = 4 (const)^2$, 
   ${\rm STr} (\Phibold \star \Phibold \star \Phibold) = 0$.

\subsection{Hermiticity conditions and reality of the NC action}

In the expansions (\ref{NCOmdef}) and (\ref{NCPhidef}) all fields are taken to be real. This 
is equivalent to the relations
 \eq
   \Ombold^\dagger = - \Gammabold_0 \Ombold \Gammabold_0, ~~ \Phibold^\dagger =  \Gammabold_0 \Phibold \Gammabold_0, ~~ \Gammabold_0 \equiv 
   \left(
\begin{array}{cc}
  \ga_0 &  0   \\
 0  &  -1   \\
\end{array}
\right)
    \en
    due to $\ga_{ab}$ and $\ga_5$ being $\ga_0$ antihermitian (i.e. $\ga_{ab}^\dagger = - \ga_0 \ga_{ab} \ga_0$ etc), while $1$, $\ga_a$ and $\ga_a \ga_5$ are $\ga_0$ -hermitian. Noting that $\Gammabold_0^2 = \onebold$, and that the $\Gammabold_0$-antihermiticity of  $\Ombold$  implies  $\Gammabold_0$-antihermiticity of $\Rbold$,  one easily proves  that the NC action is real.

\subsection{Charge conjugation invariance}

The NC action is also invariant under substitution of the fields by their charge conjugates 
 \eq
\Ombold^c \equiv -\Cbold^{-1} \Ombold^T \Cbold ~ \Rightarrow  \Rbold^c = - \Cbold^{-1} \Rbold^T \Cbold, ~~~\Phibold^c \equiv \Cbold^{-1} \Phibold^T \Cbold,~~~~
 \Cbold  \equiv 
   \left(
\begin{array}{cc}
  C &  0   \\
 0  &  1   \\
\end{array}
\right)
    \en
  and simultaneously changing $\theta$ into $-\theta$ in the $\star$-products. Indeed
\eqa
 & & S^c =  \int {\rm Str}(\Cbold^{-1} \Rbold^T \Cbold (\onebold + \unmezzo \Cbold^{-1} \Phibold^T \Cbold \Cbold^{-1} \Phibold^T \Cbold) \Cbold^{-1} \Rbold^T \Cbold \Cbold^{-1} \Phibold^T \Cbold)_{-\theta} \nonumber\\
 & & ~~~~~~=  \int {\rm Str}( \Rbold^T (\onebold + \unmezzo \Phibold^T \Phibold^T) \Rbold^T \Phibold^T )_{-\theta}\nonumber \\
& & ~~~~~~=  \int {\rm Str}( \Phibold \Rbold (\onebold + \unmezzo \Phibold\Phibold) \Rbold )^T_{\theta}  \nonumber\\
& & ~~~~~~=  \int {\rm Str}(\Rbold (\onebold + \unmezzo \Phibold\Phibold) \Rbold \Phibold )^T_{\theta}=S
\ena
\noi using ciclicity of the integral and of the supertrace, and invariance of the supertrace under matrix transposition. We have defined
 $(ABC...)_{\theta}$ to be the $\star$-(exterior) product between the forms $A,B,C...$ and 
$(ABC...)_{-\theta}$ to be the same product with opposite $\theta$. Note that for ex. $(AB)^T_{\theta}  = \pm (B^T A^T)_{-\theta} $ for $A(x),B(x)$ 
matrix valued fields (the minus sign when $A$ and $B$ are both forms of odd degree), i.e. the transposition acts only on the matrix structure of $A$ and $B$. To interchange the ordering of $A$ and $B$ as functions of $x$ one needs $\theta \rightarrow -\theta$, since
$(f \star g)_\theta = (g \star f)_{-\theta}$, as follows from the definition (\ref{starproduct}).

\sect{The geometric Seiberg-Witten map}

The results of this Section hold for any gauge group. Here we denote by $\Omhat$ the NC
gauge field, and by $\epsihat$ the NC  gauge parameter.
The Seiberg-Witten map relates  $\Omhat$
to the ordinary  $\Om$, and 
$\epsihat$ to  the ordinary  $\epsi$  so as to 
satisfy:
 \eq
\Omhat (\Om) + {\widehat{\delta}}_\epsihat
\Omhat (\Om) = \Omhat (\Om + \de_\epsi \Om)    
\label{SWcondition}
\en
 with 
  \eqa 
   & &
  \de_\epsi \Om_\mu = \part_\mu \epsi + \epsi \Om_\mu -  \Om_\mu 
      \epsi~, \\
      & &
  \deltahat_\epsihat{} \Omhat_\mu = \part_\mu \epsihat + \epsihat \star \Omhat_\mu -  \Omhat_\mu \star     
      \epsihat~.
      \ena
   
\noi  In words: the dependence of the noncommutative gauge field on the ordinary gauge field  is fixed
by requiring that ordinary gauge variations of $\Om$ inside $\Omhat(\Om)$ produce the noncommutative
gauge variation of $\Omhat$.  

Similarly noncommutative ``matter fields"  are related to the commutative
ones by requiring
 \eq
\Phihat(\Phi,\Om) + {\widehat{\delta}}_\epsihat
\Phihat (\Phi,\Om) = \Phihat (\Phi + \de_\epsi \Phi, \Om+\delta_\epsi\Omega) ~.   
\label{SWcondition1}
\en

The conditions (\ref{SWcondition}), (\ref{SWcondition1}) are satisfied if the following differential equations in the
noncommutativity parameter $\theta^{AB}$ hold \cite{SW, AC3}:
\eqa
 &  &{ \part ~~\over \part \theta^{AB}} \Omhat = {i \over 4}  \{ \Omhat^{}_{[A},  \ell^{}_{B]} \Omhat +\Rhat^{}_{B]} \}_\star ~,\label{diffeqOm} \\
 &  & { \part ~~\over \part \theta^{AB}}  \Phihat = {i \over 4}  \{ \Omhat_{[A},  \Lfat^{}_{B]} \Phihat \}_\star ~,\label{diffeqphi} \\
&  & { \part ~~\over \part \theta^{AB}}  \epsihat = {i \over 4}  \{ \Omhat_{[A},  \ell^{}_{B]} \epsihat \}_\star ~,\label{diffeqepsi} 
\ena
\noi where:\\
$\bullet$ $\Omhat_A$, $\Rhat_A$ are defined as the contraction $i_A$ along the tangent
vector $X_A$ of the exterior forms $\Omhat$, $\Rhat$, i.e. $\Omhat_A\equiv i_A\Omhat$, 
$\Rhat_A \equiv i_A \Rhat$.\\[.2em]
$\bullet$ The bracket $[A\, B]$ denotes antisymmetrization of the indices $A$ and $B$ 
with weight 1, so that for example 
$\Omhat_{[A} \Rhat_{B]} =\frac{1}{2}(\Omhat_A\Rhat_B-\Omhat_B\Rhat_A)$.
 The bracket $\{~,~\}_\star$ is the usual $\star$-anticommutator, for
example
$\{\Om_A,R_B\}_\star=\Om_A\star R_B+R_B\star\Om_A$. \\[.2em]
$\bullet$ The second differential equation holds for  fields transforming
in the adjoint representation. Notice that $\Phihat$ can  also be an
exterior form. The ``fat" Lie derivative  $\Lfat_B$ is defined by 
$\Lfat_B\equiv \ell_B+L_B\,$ where $L_B$ is the  covariant Lie derivative  along the
tangent vector $X_B$; it acts on the field $\Phihat$ as 
$$L_B  \Phihat = \ell_B \Phihat - [\Omhat_B , \Phihat]_\star\;,$$ with  
$[\Omhat_B , \Phihat]_\star=\Omhat_B \star\Phihat- \Phihat\star\Omhat_B$.
In fact the covariant Lie derivative $L_B$ can be written in Cartan form:
 \eq
  L_B = i_B D + D i_B~,
    \en
where $D$ is the covariant derivative.
\sk
The differential equations (\ref{diffeqOm})-(\ref{diffeqepsi}) hold for any abelian twist defined by arbitrary commuting vector
fields $X_A$    \cite{AC3}.  They reduce to the usual Seiberg-Witten
differential equations  \cite{SW}  in the case of a Moyal-Groenewold  twist, i.e. when $X_A\to \partial_\mu$.

\sk
We can solve these differential equations order by order in $\theta$
by expanding $\Omhat$, $\epsihat$ and $\Phihat$ in power series of $\theta$
\eqa
& & \Omhat=\Om+ \Om^1+\Om^2\;\ldots+\Om^n\ldots \\
& & \epsihat= \epsi +\epsi^1 + \epsi^2 \;\ldots + \epsi^n\ldots \\
& & \Phihat=\Phi+\Phi^1+\Phi^2\;\ldots +\Phi^n\ldots 
\ena
where the fields $\Om^n$, $\epsi^n$ and $\Phi^n$ are homogeneous polynomials in
$\theta$ of order $n$. 
By multiplying the differential equations by $\theta^{AB}$ and using the
identities
$\theta^{AB}\frac{\partial}{\partial
  \theta^{AB}}\Om^{n+1}=(n+1)\Om^{n+1}$ and similar for $\epsi^{n+1}$ and $\Phi^{n+1}$,
we obtain the recursive relations
\eqa
& & \Om^{n+1} = \frac{i\,\theta^{AB}}{4(n+1)} \{\Omhat_A,  \ell_B \Omhat +\Rhat_B\}_\star^{n} \label{Omn+1} ~,\\
& & \Phi^{n+1}  =\frac{i\,\theta^{AB}}{4(n+1)}  \{\Omhat_A,  \Lfat_B \Phihat \}_\star^{n} \label{phin+1}~, \\
& & \epsi^{\;n+1}  =\frac{i\,\theta^{AB}}{4(n+1)}  \{\Omhat_A,  \ell_B
\epsihat \}_\star^{n} \label{epsin+1}~,
\ena
where for any field $P$ (also composite, as for ex.  $\{\Omhat_A,
\Lfat_B \Phihat \}_\star$),
$P^n$ denotes its component of
order $n$ in $\theta$. These recursion relations reduce to the ones found in ref. \cite{Ulker}
in the special case of a Moyal twist.

In the following we omit the hat 
denoting noncommutative fields,  the $\star$ and $\wedge_\star$
products, and simply write $\{~,~\}$, $[~,~]$  for
$\{~,~\}_\star$, $[~,~]_\star$. 

If $P$ and $Q$ are forms in the adjoint representation of the gauge group (i.e. if $\de_\epsi P = - P \epsi + \epsi  P$ etc.) the following recursion relation for the product $PQ$  holds \cite{ACD2}:
\eq
(PQ)^{n+1}=\frac{i\,\theta^{AB}}{4(n+1)}\Big(\{\Om_A,\Lfat_B(P Q)\}
+2L_A P \, L_B Q
\Big)^{n}~\label{RRCF}.
\en
Some other useful identities are \cite{ACD2}:
 \eqa
 & & \theta^{AB} L_A L_B P = - \unmezzo \theta^{AB} \{R_{AB},P \} \\
 & & \theta^{AB} \Lfat_A \Om_B  =  \theta^{AB} R_{AB} \\
 & & \theta^{AB}\int Tr\Big( \{\Om_A,I\!\!L_B(PQ)\} +2L_AP\,L_B
Q\,\Big)=\,\theta^{AB}\int Tr \Big(\{R_{AB}, P\}Q\Big)~
 \ena
 \noi where $R_{AB} \equiv i_B i_A R $. Finally, using (\ref{RRCF}) one can find the recursion relation for the curvature:
 \eq
 R^{n+1}
   =\frac{i\,\theta^{AB}}{4(n+1)}\Big(\{\Om_A,\Lfat_B R\} -
[R_A,R_B]\Big)^{n} \label{Rn+1}
 \en
 Some basic formulae of Cartan calculus, used in deriving the above identities,  are listed in Appendix A.
 \sk
We list below the first order corrections to the classical
$OSp(1|4)$ fields and curvatures, obtained by using the general recursion formulas (\ref{Omn+1}), (\ref{phin+1}) and (\ref{Rn+1}) for $n=0$. On the right-hand sides all products are ordinary exterior products, and all fields are classical.

\subsection{$OSp(1|4)$ fields and curvatures at first order in $\theta$}
\noi {\bf $\Ombold$ connection components}
\eqa
& & \Om^1={i \over 4}\theta^{AB} (\{ \Om_A,\ell_B \Om + R_B \} + \psi_A \ell_B \psibar
+ \ell_B \psi \psibar_A + \psi_A \Sigmabar_B + \Sigma_B \psibar_A ) \\
 & & \psi^1 = {i \over 4}\theta^{AB} (\Om_A \ell_B \psi + \ell_B \Om \psi_A 
 + \Om_A \Sigma_B +R_B \psi_A) \\
 & & w^1 = {i \over 4}\theta^{AB} (\psibar_A \ell_B \psi + \ell_B \psibar \psi_A + \psibar_A \Sigma_B+ \Sigmabar_B \psi_A)
 \ena

\noi {\bf $\Phibold$ field components}
\eqa
 & & \Phi^1 = -{1 \over 4}\theta^{AB} (\{ \Om_A , \ga_5 \Om_B - \Om_B \ga_5 \} +
  \psi_A \psibar_B \ga_5) \\
& & \zeta^1 = - {1 \over 4}\theta^{AB}(2 \Om_A \ga_5 \psi_B + \ga_5 \Om_B \psi_A) \\
 & & \pi^1 = -{1 \over 2}\theta^{AB} (\psibar_a \ga_5 \psi_B)
 \ena

\noi {\bf $\Rbold$ curvature components}
 \eqa 
 & & R^1 = {i \over 4}\theta^{AB} (\{\Om_A , \Lfat_B R \}
 - [R_A,R_B]+ \{\Om_A,-\psi_B \Sigmabar + \sigma \psibar_B \}
  \nonumber\\
 & & ~~~~~~+ \psi_A \Lfat_B \Sigmabar + \Lfat_B \Sigma \psibar_A - \psi_A \psibar_B R + R \psi_B \psibar_A - 2 \Sigma_A \Sigmabar_B)\\
 & & \Sigma^1 = {i \over 4}\theta^{AB} (\Om_A (\Lfat_B \Sigma +
  R \psi_B) - \psi_A (\psibar_B \Sigma + \Sigmabar \psi_B )+
 (\Sigma \psibar_B - \psi_B \Sigmabar) \psi_A \nonumber\\
 & &~~~~~~ + \Lfat_B R \psi_A - 2 R_A \Sigma_B) \\
& & r^1={i \over 4}\theta^{AB}(\psibar_A \Lfat_B \Sigma + \Lfat_B \Sigmabar \psi_A + 2 \psibar_A R \psi_B - 2 \Sigmabar_A \Sigma_B)
 \ena

\sect{The extended $OSp(1|4)$ supergravity action}

We now discuss the $\theta$ expansion of the NC action (\ref{action4}) , where the NC supermatrix fields 
$\Ombold$ and $\Phibold$ have been substituted by their SW expansion in terms of the classical fields. 

\subsection{The action is even in $\theta$}

We first note that the SW map is such that:
 \eqa
 & &  \Ombold_\theta^c \equiv -\Cbold^{-1} \Ombold^T_\theta \Cbold =  \Ombold_{-\theta}, ~~~ \Rightarrow  \Rbold_\theta^c = -  \Cbold^{-1} \Rbold^T_\theta \Cbold =  \Rbold_{-\theta} \label{CCrel1} \\
& & \Phibold_\theta^c \equiv \Cbold^{-1} \Phibold^T_\theta \Cbold =  \Phibold_{-\theta}
\label{CCrel2} \ena

 \noi where the $\theta$ dependence is explicitly indicated as a subscript. The proof by induction, using (\ref{Omn+1}) and (\ref{phin+1}),  is straightforward. Suppose that relations (\ref{CCrel1})  hold up to order 
$\theta^n$. Then 
 \eqa
 & &-  \Cbold^{-1} (\Ombold^T)^{n+1}_\theta \Cbold = \nonumber \\
  & &~~~~~~ = {-i \theta^{AB} \over 4(n+1)} (\Cbold^{-1} \Ombold^T_\theta \Cbold \Cbold^{-1} (\ell_B \Ombold + \Rbold)^T_\theta \Cbold + \Cbold^{-1}(\ell_B \Ombold + \Rbold)^T_\theta \Cbold \Cbold^{-1}\Ombold^T_\theta \Cbold )^n
\nonumber \\
 & & ~~~~~~ = {-i \theta^{AB} \over 4(n+1)} (\Ombold_{-\theta} (\ell_B \Ombold + \Rbold)_{-\theta} + (\ell_B \Ombold + \Rbold)_{-\theta} \Ombold_{-\theta})^n  \nonumber \\
 & & ~~~~~~ =  \Ombold_{-\theta}^{n+1}
 \ena
 Similarly one proves (\ref{CCrel2}). Exploiting now the invariance of the NC action $S$ under charge conjugation, proved in Section 4, and using the relations (\ref{CCrel1}) and (\ref{CCrel2}) one finally finds:
 \eq
 S_\theta = S^c_\theta = S_{-\theta}
 \en
i.e. the NC action $S$ {\it is even } in $\theta$. Therefore the $\theta$-expansion of $S$ has the form:
 \eq
 S = S^0 + S^2 + S^4 + \cdots
 \en
and the first nonvanishing correction to the classical action $S^0$ is at order $\theta^2$. 

Note that the relations (\ref{CCrel1}) and (\ref{CCrel2}) imply the following conditions on the NC component fields:
  \eqa
 & & \om^{ab}_\theta=\om^{ab}_{-\theta},~~ V^{a}_\theta=V^{a}_{-\theta}, ~~C \psibar^T_\theta = \psi_{-\theta}\\
 & &  \om_\theta=- \om_{-\theta}, ~~ \omtilde_\theta= - \omtilde_{-\theta}, ~~ \Vtilde^{a}_\theta= - \Vtilde^{a}_{-\theta} 
 \ena
 \noi and
  \eqa
 & &  \pi_\theta= \pi_{-\theta}, ~~ \phi_\theta=  \phi_{-\theta}, ~~ \phitilde^{a}_\theta=  \phitilde^{a}_{-\theta},~~C \zetabar^T_\theta = \zeta_{-\theta} \\
  & & \phi^{ab}_\theta= - \phi^{ab}_{-\theta},~~ \phi^{a}_\theta= - \phi^{a}_{-\theta} 
 \ena
 \noi where the $\theta$ dependence of the NC fields is indicated with a subscript. Thus in the limit $\theta \rightarrow 0$
 we see that only $\om^{ab}$, $V^a$ and $\psi$ survive in $\Ombold$, and only $\pi$, $\phi$, $\phitilde^a$ and $\zeta$ survive
 in $\Phibold$, in agreement with the classical fields in (\ref{Omdef}) and (\ref{Phidef}). Finally we recall that $C \psibar^T_\theta = \psi_{-\theta}$
(and similar for $\zeta$) is the noncommutative definition for a Majorana spinor \cite{AC1,AC2}, consistent with the NC 
gauge transformations and reducing to the usual definition for $\theta=0$.

\subsection{The action at order $\theta^2$}

We can compute the $\theta^2$ correction with the help of the recursion relations (\ref{RRCF})  for composite fields, and the identities
at the end of section 5. The result reads:
 \eq
 S^2 = S^2_{RR\Phi} +  S^2_{R \Phi \Phi R \Phi}
 \en
\noi with
\eqa
 & & S^2_{RR\Phi} = - {1 \over 16} \theta^{AB} \theta^{CD}
  \int {\rm STr} (R_{AB} R_{CD} RR\Phi + \unmezzo \{R_{CD},RR \} R_{AB} \Phi-2 R_{AC}R_{BD} \{RR,\Phi\} \nonumber \\ & & ~~+ \{R_{AB},L_CR\} L_D R \Phi + \{R_{AB}, \Phi \} L_C R L_D R \Phi + 2 \{R_{AC},L_D R \} [L_B R ,\Phi] \nonumber \\ & & ~~ - \{R_{CD},R_A R_B \} R \Phi - \{R_{CD} , R \}R_A R_B \Phi - R_{AB} \{R_C R_D , R \}\{ \Phi, R_{AB}\} \nonumber \\
 & & ~~+ R_{AB} L_C (RR) L_D \Phi + RR L_C R_{AB} L_D \Phi - 2 L_A (R_C R_D) \{L_B R . \Phi \}+ 2 L_A R (L_C L_B R ) L_D \Phi \nonumber \\
& &  ~~ - L_C R_A L_D R_B R \Phi - 2 R L_C (R_A R_B) L_D \Phi - 2 R 
L_C R_A L_D R_B \Phi \nonumber \\
 & & ~~ +2 i_A (R_C R_D) ( \{R_B,R \Phi \}+ [R_B, \Phi R]) + 2 R_A R_B L_C R L_D \Phi + 4 R_A R_B R_C R_D \Phi)
 \ena
\eqa
 & & S^2_{R \Phi \Phi R \Phi} =  - {1 \over 16} \theta^{AB} \theta^{CD} \int {\rm STr} (( \unmezzo \{R_{CD}, \{R_{AB},R \Phi \Phi \} \} - \{ R_{CD} , \{ R_A R_B \Phi , \Phi \} \} \nonumber \\ 
 & & ~ + \{ R_{CD}, L_A R \{ \Phi , L_B \Phi \} \} + \{ R_{CD} , R L_A \Phi L_B \Phi \} + \{ R_{CD} , \Phi L_A R L_B \Phi \}  - \{ \{ R_{AC}, R_{BD} \}, R \Phi\Phi \}  \nonumber \\
 & & ~ + [L_C R_{AB} , L_D (R \Phi\Phi)]+ \{ R_{AB}, L_C R L_D(\Phi\Phi)- R_C R_D \Phi\Phi + R L_C \Phi L_D \Phi \} \nonumber \\
 & & ~- [L_C (R_A R_B ) , L_D \Phi] - \{L_C R_A L_D R_B \Phi, \Phi \} + \{ [ i_A (R_C R_D ) , R_B \Phi] , \Phi \}  + L_C L_A (R \Phi) L_D L_B \Phi   \nonumber \\
 & & ~ + ((L_CL_A R)L_D \Phi + \{R_{AC}, L_D R \} \Phi - L_A(R_C R_D\Phi))L_B \Phi + L_A (R \Phi )\{R_{BC}, L_D \Phi \}  \nonumber \\
 & & ~+( L_CRL_DL_A \Phi L_B \Phi + R \{R_{AC},L_D \Phi \}  L_B \Phi + [L_C(L_ARL_B\Phi),L_D \Phi]  \nonumber \\
 & & ~ + \{ L_C L_A R L_D L_B \Phi + \{ R_{AC} , L_D R \} L_B \Phi - L_A (R_CR_D)L_B \Phi + L_AR \{R_{BC}, L_D \Phi \} , \Phi \} ) R \Phi \nonumber \\
 & & ~+2 (\unmezzo \{R_{AB}, R\Phi\Phi \} - \{R_AR_B \Phi , \Phi \} + L_A(R \Phi) L_B \Phi+ \{L_AR LL_B \Phi, \Phi \})(L_CR L_D\Phi - R_C R_D \Phi) ) \nonumber \\
 \ena
Here all products are ordinary exterior products between classical fields. These corrections to the classical $OSp(1|4)$ action are invariant under local (ordinary) $OSp(1|4)$ gauge variations, as is manifest since all quantities appearing in $S^2$ are gauge covariant, and transform in the adjoint (i.e. as commutators with the gauge parameter). The SW map is designed  to ensure this invariance: to find explicitly gauge invariant corrections, order by order in $\theta$, is a powerful check on the computations.               

To recover the usual $N=1$, $D=4$ AdS supergravity (without auxiliary fields) in the $\theta \rightarrow 0$  limit one still needs to break $OSp(1|4)$ to its Lorentz subroup. This is done exactly as in the classical case, by choosing the gauge where $\Phibold$ becomes the constant supermatrix $\Gammabold$ defined in (\ref{Gammabold})
(the constrained auxiliary fields  take constant values). This gauge breaks translations and supersymmetry. We have seen how supersymmetry can be uncovered in the classical ($\theta=0$) gauge fixed action. The question whether a hidden supersymmetry is present also in the gauge fixed extended  ($\theta \neq 0$) action is left to future investigations.

\sect{Conclusions}

The fascinating idea that (super)gravity has some kind of conformal phase, before breaking occurs and dimensionful  constants emerge is rather old and the  $OSp(4|1)$ actions we have been discussing are part of this idea. 

The result we have presented here is a noncommutative extension of $OSp(4|1)$ supergravity, the novelty being on one
side a $D=4$ {\it supergravity} action $S$ invariant under local $\star$-supersymmetry (part of the supergroup noncommutative gauge symmetry), and on the other side explicit invariance of $S$ under diffeomorphisms, thanks to a geometrical formulation of abelian twists.  Previous works have addressed noncommutative extensions of Mac Dowell-Mansouri gravity actions, but without treating their supersymmetric versions. 

We have then used
a generalization of the Seiberg-Witten map (adapted to abelian twists and suitably ``geometrized"), obtaining a higher derivative $D=4$ supergravity, with constrained  auxiliary fields, invariant under the {\it usual} gauge transformations of the whole supergroup $OSp(1|4)$.  Recursion formulae for the SW higher order corrections have been applied to 
compute the  $\theta^2$ correction to the classical $OSp(1|4)$ action.

In short, noncommutativity has been used as a guide to construct an extended, locally supersymmetric  higher derivative theory with the same symmetries of its classical $\theta \rightarrow 0$ limit.   

\sk\sk

\noi {\bf Acknowledgements}
 \sk
 \noi We thank Paolo Aschieri for useful comments on the manuscript. 

\appendix

\sect{Cartan formulae}
The usual Cartan calculus formulae simplify if we consider commuting
vector fields $X_A$, and read 
\eqa
& & \ell_A=i_Ad+di_A ~~,~~~~~~~~~~~~~~~~~~~~~~~~~~~~L_A=i_AD+D i_A\\
& & [\ell_A,\ell_B]=0~~,~~~~~~~~~~~~~~~~~~~~~~~~~~~~~~~~~[L_A,L_B]=i_Ai_B R\\
& & [\ell_A,i_B]=0~~,~~~~~~~~~~~~~~~~~~~~~~~~~~~~~~~~~[L_A,i_B]=0\\
& & i_Ai_B+i_Bi_A=0~~,~~~~d\circ d=0~~,~~~~~~~~D\circ D= R
\ena

\sect{Gamma matrices in $D=4$}

We summarize in this Appendix our gamma matrix conventions in $D=4$.

\eqa
& & \eta_{ab} =(1,-1,-1,-1),~~~\{\ga_a,\ga_b\}=2 \eta_{ab},~~~[\ga_a,\ga_b]=2 \ga_{ab}, \\
& & \ga_5 \equiv i \ga_0\ga_1\ga_2\ga_3,~~~\ga_5 \ga_5  = 1,~~~\epsi_{0123} = - \epsi^{0123}=1, \\
& & \ga_a^\dagger = \ga_0 \ga_a \ga_0, ~~~\ga_5^\dagger = \ga_5 \\
& & \ga_a^T = - C \ga_a C^{-1},~~~\ga_5^T = C \ga_5 C^{-1}, ~~~C^2 =-1,~~~C^\dagger=C^T =-C
\ena

\subsection{Useful identities}

\eqa
 & &\ga_a\ga_b= \ga_{ab}+\eta_{ab}\\
 & & \ga_{ab} \ga_5 = {i \over 2} \epsilon_{abcd} \ga^{cd}\\
 & &\ga_{ab} \ga_c=\eta_{bc} \ga_a - \eta_{ac} \ga_b -i \epsi_{abcd}\ga_5 \ga^d\\
 & &\ga_c \ga_{ab} = \eta_{ac} \ga_b - \eta_{bc} \ga_a -i \epsi_{abcd}\ga_5 \ga^d\\
 & &\ga_a\ga_b\ga_c= \eta_{ab}\ga_c + \eta_{bc} \ga_a - \eta_{ac} \ga_b -i \epsi_{abcd}\ga_5 \ga^d\\
 & &\ga^{ab} \ga_{cd} = -i \epsi^{ab}_{~~cd}\ga_5 - 4 \de^{[a}_{[c} \ga^{b]}_{~~d]} - 2 \de^{ab}_{cd}\\
& & Tr(\ga_a \ga^{bc} \ga_d)= 8~ \de^{bc}_{ad} \\
& & Tr(\ga_5 \ga_a \ga_{bc} \ga_d) = -4 i~\epsi_{abcd} 
 \ena
\sk
\noi where
$\delta^{ab}_{cd} \equiv \frac{1}{2}(\delta^a_c\delta^b_d-\delta^b_c\delta^a_d)$, $\delta^{rse}_{abc} \equiv  {1 \over 3!} (\de^r_a \de^s_b \de^e_c$ + 5 terms), 
and indices antisymmetrization in square brackets has total weight $1$.

 \subsection{Charge conjugation and Majorana condition}

\eqa
 & &   {\rm Dirac~ conjugate~~} \psibar \equiv \psi^\dagger
 \ga_0\\
 & &  {\rm Charge~ conjugate~spinor~~} \psi^C = C (\psibar)^T  \\
 & & {\rm Majorana~ spinor~~} \psi^C = \psi~~\Rightarrow \psibar =
 \psi^T C
 \ena

\end{document}